\documentclass[zpreprint,zbstepj]{zeus_paper}
\usepackage[english]{babel}
\chardef\usc=95
\chardef\til=126
\catcode`\@=11 
\DeclareRobustCommand\xdotspace{\futurelet\@let@token\@xdotspace}
\def\@xdotspace{%
  \ifx\@let@token.\else
  \ifx\@let@token\bgroup.\else
  \ifx\@let@token\egroup.\else
  \ifx\@let@token\/.\else
  \ifx\@let@token\ .\else
  \ifx\@let@token~.\else
  \ifx\@let@token!.\else
  \ifx\@let@token,.\else
  \ifx\@let@token:.\else
  \ifx\@let@token;.\else
  \ifx\@let@token?.\else
  \ifx\@let@token/.\else
  \ifx\@let@token'.\else
  \ifx\@let@token).\else
  \ifx\@let@token-.\else
  \ifx\@let@token\@xobeysp.\else
  \ifx\@let@token\space.\else
  \ifx\@let@token\@sptoken.\else
   .\space
   \fi\fi\fi\fi\fi\fi\fi\fi\fi\fi\fi\fi\fi\fi\fi\fi\fi\fi}
\catcode`\@=12 

\newcommand{\stru}[2]{%
   \relax\ifmmode\hbox{\vrule height#1 depth#2 width0pt}%
   \else\vrule height#1 depth#2 width0pt\fi}

\newcommand{\Ronum}[1]{\uppercase\expandafter{\romannumeral#1}}
\newcommand{\ronum}[1]{\expandafter{\romannumeral#1}}
\DeclareRobustCommand{\LaTeXZ}{%
  \LaTeX\kern-.05em4\kern-.1em
  {\raisebox{-0.2ex}{$\scriptstyle\text{ZEUS}$}}\xspace}

\DeclareMathAlphabet{\mathbf}{OT1}{cmr}{bx}{sl}
\newcommand{\eVdist}{\kern-0.06667em}

\newcommand{\Gev}{{\text{Ge}\eVdist\text{V\/}}}

\newcommand{\gev}{{\,\text{Ge}\eVdist\text{V\/}}}

\newcommand{\Tesla}{\,\text{T}}

\newcommand{\slashfrac}[2]{%
  \raisebox{0.5ex}{\ensuremath #1}\kern-0.12em/\kern-0.08em
  \raisebox{-.8ex}{\ensuremath #2}}

\newcommand{\sqr}[3]{%
    {\vcenter{\hrule height.#3ex\hbox{\vrule width.#2ex height#1ex
     \kern#1ex\vrule width.#3ex}\hrule height.#2ex}}}

\catcode`\@=11 
\newcommand{\parenbar}{\mathpalette\p@renb@r}
\def\p@renb@r#1#2{\vbox{%
  \ifx#1\scriptscriptstyle \dimen@.7em\dimen@ii.2em\else
  \ifx#1\scriptstyle \dimen@.8em\dimen@ii.25em\else
  \dimen@1em\dimen@ii.4em\fi\fi \offinterlineskip
  \ialign{\hfill##\hfill\cr
    \vbox{\hrule width\dimen@ii}\cr
    \noalign{\vskip-.3ex}%
    \hbox to\dimen@{$\mathchar300\hfil\mathchar301$}\cr
    \noalign{\vskip-.3ex}%
    $#1#2$\cr}}}
\catcode`\@=12 

\newcommand{\MSbar}{\hbox{$\overline{\rm MS}$}\xspace}

\newcommand{\IP}{{\rm I$\kern-0.01667em$P}\xspace}

\mathchardef\qsm=63
\mathchardef\pls=43
\mathchardef\mns=512
\mathchardef\plm=518
\mathchardef\eql=61
\mathchardef\smallleft=300
\mathchardef\smallright=301
\mathchardef\les=316
\mathchardef\gre=318
\mathchardef\leq=532
\mathchardef\grq=533

\catcode`\@=11 
\newcounter{pict@width}
\newcounter{pict@height}
\newlength{\pict@scale}
\newcommand{\psfigadd}[4]{%
\setcounter{pict@width}{1*\ratio{#2+\pict@scale/2}{\pict@scale}}
\setcounter{pict@height}{1*\ratio{#3+\pict@scale/2}{\pict@scale}}
\setlength{\unitlength}{\pict@scale}
\hbox to #2{\hspace{-\fill}\begin{picture}(\thepict@width,\thepict@height)
\put(0,0){\psfig{figure=#1,width=#2,height=#3,clip=}}
\SetScale{0.283466457}
\SetWidth{1.763889}
{#4}
\end{picture}}
}
\newcounter{pict@widthfst}
\newcounter{pict@widthscd}
\newcounter{pict@widthtot}
\newcommand{\psfigaddtwo}[7]{%
\setcounter{pict@widthfst}{1*\ratio{#2+\pict@scale/2}{\pict@scale}}
\setcounter{pict@widthscd}{1*\ratio{#2+#4+\pict@scale/2}{\pict@scale}}
\setcounter{pict@widthtot}{1*\ratio{#2+#4+#6+\pict@scale/2}{\pict@scale}}
\setcounter{pict@height}{1*\ratio{#3+\pict@scale/2}{\pict@scale}}
\setlength{\unitlength}{\pict@scale}
\hbox{\hspace{-\fill}\begin{picture}(\thepict@widthtot,\thepict@height)
\put(0,0){\psfig{figure=#1,width=#2,height=#3,clip=}}
\put(\thepict@widthscd,0){\psfig{figure=#5,width=#6,height=#3,clip=}}
\SetScale{0.283466457}
\SetWidth{1.763889}
{#7}
\end{picture}}
}
\newcommand{\psfigror}[4]{%
\setcounter{pict@width}{1*\ratio{#2+\pict@scale/2}{\pict@scale}}
\setcounter{pict@height}{1*\ratio{#3+\pict@scale/2}{\pict@scale}}
\setlength{\unitlength}{\pict@scale}
\hbox{\begin{picture}(\thepict@width,\thepict@height)
\put(0,\thepict@height){\psfig{figure=#1,width=#3,height=#2,clip=,angle=270}}
\SetScale{0.283466457}
\SetWidth{1.763889}
{#4}
\end{picture}}
}
\newcommand{\psfigrol}[4]{%
\setcounter{pict@width}{1*\ratio{#2+\pict@scale/2}{\pict@scale}}
\setcounter{pict@height}{1*\ratio{#3+\pict@scale/2}{\pict@scale}}
\setlength{\unitlength}{\pict@scale}
\hbox{\begin{picture}(\thepict@width,\thepict@height)
\put(0,0){\psfig{figure=#1,width=#3,height=#2,clip=,angle=90}}
\SetScale{0.283466457}
\SetWidth{1.763889}
{#4}
\end{picture}}
}
\catcode`\@=12 
\newlength\listtextwidth

\catcode`\@=11 
\newlength{\@tabfninsert}
\newlength{\@tabfnwidth}
\newcommand{\tabfootnote}[2]{%
  \setlength{\@tabfninsert}{0.8em}
  \setlength{\@tabfnwidth}{\textwidth}
  \addtolength{\@tabfnwidth}{-\@tabfninsert}
  \addtolength{\@tabfnwidth}{-0.4em}
  \noindent\makebox[\@tabfninsert][r]{\footnotesize$^{#1}$\hfil}\hfill%
  \parbox[t]{\@tabfnwidth}{\footnotesize #2\hfill}}
\catcode`\@=12 

\begin{document}

\prepnum{DESY-05-019}

\title{
Multijet production in neutral current deep inelastic scattering at HERA and 
determination of $\alpha_s$ 
}  

\author{ZEUS Collaboration}
 
\date{January 2005}

\abstract{
Multijet production rates in neutral current deep inelastic 
scattering have been measured in the range of exchanged boson
virtualities $10 < Q^2 < 5000\,\hbox{GeV}^2$. The data were taken at
the $ep$ collider HERA with centre-of-mass energy  $\sqrt{s} =
318\,\hbox{GeV}$ using the ZEUS detector and correspond to an
integrated luminosity of $82.2\,\hbox{pb}^{-1}$. Jets were identified in the Breit frame
using the $k_T$ cluster algorithm in the longitudinally invariant inclusive mode. 
Measurements of differential dijet and trijet cross sections are presented as
functions of jet transverse energy ($E_{T,B}^{\rm jet}$), pseudorapidity
($\eta_{\rm LAB}^{\rm jet}$) and $Q^2$ with $E_{T,B}^{\rm jet} > 5\,\hbox{GeV}$ and $-1 <
\eta_{\rm LAB}^{\rm jet} < 2.5$. Next-to-leading-order QCD calculations describe the data well.
The value of the strong coupling constant $\alpha_s(M_Z)$, determined from
the ratio of the trijet to dijet cross sections, is
$\alpha_s(M_Z) = 0.1179 \pm 0.0013~\rm(stat.)^{+0.0028}_{-0.0046}~\rm(exp.)^{+0.0064}_{-0.0046}~\rm(th.)$. 
}

\makezeustitle

\pagenumbering{Roman}                                                                              
                                                   %
\begin{center}                                                                                     
{                      \Large  The ZEUS Collaboration              }                               
\end{center}                                                                                       
  S.~Chekanov,                                                                                     
  M.~Derrick,                                                                                      
  S.~Magill,                                                                                       
  S.~Miglioranzi$^{   1}$,                                                                         
  B.~Musgrave,                                                                                     
  \mbox{J.~Repond},                                                                                
  R.~Yoshida\\                                                                                     
 {\it Argonne National Laboratory, Argonne, Illinois 60439-4815}, USA~$^{n}$                       
\par \filbreak                                                                                     
  M.C.K.~Mattingly \\                                                                              
 {\it Andrews University, Berrien Springs, Michigan 49104-0380}, USA                               
\par \filbreak                                                                                     
  N.~Pavel, A.G.~Yag\"ues Molina \\                                                                
  {\it Institut f\"ur Physik der Humboldt-Universit\"at zu Berlin,                                 
           Berlin, Germany}
% ...
\newpage                                                                                           
$^{\    1}$ also affiliated with University College London, UK \\                                  
% ...
                                                           %
                                                           %
\newpage   
                                                           %
                                                           %
\begin{tabular}[h]{rp{14cm}}                                                                       
$^{a}$ &  supported by the Natural Sciences and Engineering Research Council of Canada (NSERC) \\  
$^{b}$ &  supported by the German Federal Ministry for Education and Research (BMBF), under        
          contract numbers HZ1GUA 2, HZ1GUB 0, HZ1PDA 5, HZ1VFA 5\\                                
$^{c}$ &  supported in part by the MINERVA Gesellschaft f\"ur Forschung GmbH, the Israel Science   
          Foundation (grant no. 293/02-11.2), the U.S.-Israel Binational Science Foundation and    
          the Benozyio Center for High Energy Physics\\                                            
$^{d}$ &  supported by the German-Israeli Foundation and the Israel Science Foundation\\           
$^{e}$ &  supported by the Italian National Institute for Nuclear Physics (INFN) \\                
$^{f}$ &  supported by the Japanese Ministry of Education, Culture, Sports, Science and Technology 
          (MEXT) and its grants for Scientific Research\\                                          
$^{g}$ &  supported by the Korean Ministry of Education and Korea Science and Engineering          
          Foundation\\                                                                             
$^{h}$ &  supported by the Netherlands Foundation for Research on Matter (FOM)\\                   
$^{i}$ &  supported by the Polish State Committee for Scientific Research, grant no.               
          620/E-77/SPB/DESY/P-03/DZ 117/2003-2005 and grant no. 1P03B07427/2004-2006\\             
$^{j}$ &  partially supported by the German Federal Ministry for Education and Research (BMBF)\\   
$^{k}$ &  supported by RF Presidential grant N 1685.2003.2 for the leading scientific schools and  
          by the Russian Ministry of Education and Science through its grant for Scientific        
          Research on High Energy Physics\\                                                        
$^{l}$ &  supported by the Spanish Ministry of Education and Science through funds provided by     
          CICYT\\                                                                                  
$^{m}$ &  supported by the Particle Physics and Astronomy Research Council, UK\\                   
$^{n}$ &  supported by the US Department of Energy\\                                               
$^{o}$ &  supported by the US National Science Foundation\\                                        
$^{p}$ &  supported by the Polish Ministry of Scientific Research and Information Technology,      
          grant no. 112/E-356/SPUB/DESY/P-03/DZ 116/2003-2005 and 1 P03B 065 27\\                  
$^{q}$ &  supported by FNRS and its associated funds (IISN and FRIA) and by an Inter-University    
          Attraction Poles Programme subsidised by the Belgian Federal Science Policy Office\\     
\end{tabular}                                                                                      
                                                           %
                                                           %
                                                                                      
\pagenumbering{arabic} 
\pagestyle{plain}
\raggedbottom
\section{Introduction}
\label{intro}

Measurements of multijet production from initial-state hadrons and leptons have been carried
out previously in collisions at the SPS \cite{pl:b158:494,*zfp:c30:341}, the ISR 
\cite{np:b303:569,*zfp:c32:317}, the TEVATRON \cite{prl:86:1955,*pr:d53:6000}
and at LEP \cite{proc:moriond:1996:223,*npps:74:44} as well as in photoproduction \cite{pl:b443:394} and deep inelastic 
scattering (DIS) \cite{epj:c6:575} at HERA.
Multijet production in DIS at HERA has been used to test the predictions of perturbative QCD (pQCD) 
calculations 
over a large range of four-momentum transfer squared, $Q^2$ \cite{pl:b515:17}.
Recently, the ZEUS and H1 collaborations have determined the strong coupling
constant $\alpha_s$ from a variety of measurements of jet production and jet properties
in both DIS \cite{epj:c6:575,pl:b363:201,pl:b507:70,pl:b547:164,pl:b558:41,epj:c19:289,epj:c21:33,np:b700:3}
and photoproduction \cite{pl:b560:7}.

At leading order (LO) in $\alpha_s$, dijet production in neutral current DIS
proceeds via the boson-gluon-fusion (BGF, $V^{*} g \rightarrow q\bar{q}$ with $V=\gamma$, $Z^0$) 
and QCD-Compton (QCDC, $V^{*} q \rightarrow qg$) processes.
Events with three jets can be seen as dijet processes with 
an additional gluon radiation or splitting of a gluon
into a quark-antiquark pair and are directly sensitive to
$\mathcal{O}(\alpha_s^{2})$ QCD effects.
The higher sensitivity to $\alpha_s$ and the large number of degrees of freedom of the trijet final state 
allow detailed testing of QCD predictions.

In the present analysis, the differential cross sections for the trijet production have been measured with high statistical precision. 
Measurements of the inclusive trijet cross section as a function 
of $Q^2$ and the jet transverse energy, $E_{T,B}^{\rm jet}$, in the Breit frame and the jet 
pseudorapidity, $\eta_{\rm LAB}^{\rm jet}$, in the laboratory frame are presented.
Predictions of pQCD at next-to-leading order (NLO) are compared to the measurements. 
In addition, the analysis includes the first $\alpha_s$ determination using the cross-section ratio 
of trijet to dijet production, $R_{3/2}$, at HERA. In this ratio, correlated
experimental and 
theoretical uncertainties cancel, allowing for an extension 
of the measurement to low $Q^2$.

\section{Experimental set-up}
\label{exp}
The data used in this analysis were collected during the 1998-2000 running
period, when HERA operated with protons of energy
$E_p=920$~GeV and electrons or positrons\footnote{In the
following, the term ``electron'' denotes generically both the
electron ($e^-$) and the positron ($e^+$).}
of energy $E_e=27.5$~GeV, and correspond to an integrated luminosity
of $82.2\pm 1.9$~pb$^{-1}$. A detailed description of the ZEUS detector
can be found elsewhere~\cite{pl:b293:465,zeus:1993:bluebook}. A brief
outline of the components that are most relevant for this analysis is
given below.

Charged particles are measured in the central tracking detector
(CTD)~\cite{nim:a279:290,*npps:b32:181,*nim:a338:254}, which operates in a
magnetic field of $1.43\Tesla$ provided by a thin superconducting
solenoid. The CTD consists of 72~cylindrical drift chamber
layers, organised in nine superlayers covering the
polar-angle\footnote{The 
ZEUS coordinate system is a right-handed Cartesian system, with the $Z$ axis 
pointing in the proton beam direction, referred to as the 
``forward direction'', and the $X$ axis pointing left towards the centre of 
HERA. The coordinate origin is at the nominal interaction point.} 
region \mbox{$15^\circ<\theta<164^\circ$}. The transverse momentum
resolution for full-length tracks can be parameterised as
$\sigma(p_T)/p_T=0.0058p_T\oplus0.0065\oplus0.0014/p_T$, with $p_T$ in
$\Gev$. The tracking system was used to measure the interaction vertex
with a typical resolution along (transverse to) the beam direction of
0.4~(0.1)~cm and also to cross-check the energy scale of the calorimeter.

The high-resolution uranium-scintillator calorimeter
(CAL)~\cite{nim:a309:77,*nim:a309:101,*nim:a321:356,*nim:a336:23} covers
$99.7\%$ of the total solid angle and consists of three parts: the
forward (FCAL), the barrel (BCAL) and the rear (RCAL) calorimeters. Each
part is subdivided transversely into towers and longitudinally into one
electromagnetic section (EMC) and either one (in RCAL) or two (in BCAL and
FCAL) hadronic sections (HAC). The smallest subdivision of the calorimeter
is called a cell. Under test-beam conditions, the CAL single-particle
relative energy resolutions were $\sigma(E)/E=0.18/\sqrt{E}$ for
electrons and $\sigma(E)/E=0.35/\sqrt{E}$ for hadrons, with $E$ in GeV.

The luminosity was measured from the rate of the bremsstrahlung process
$ep\rightarrow e\gamma p$. The resulting small angle energetic photons
were measured by the luminosity
monitor~\cite{desy-92-066,*zfp:c63:391,*acpp:b32:2025}, a
lead-scintillator calorimeter placed in the HERA tunnel at $Z=-107$ m.

\section{Kinematics and event selection}
\label{ks}
A three-level trigger system was used to select events online \cite{zeus:1993:bluebook,zfp:c74:207,thesis:krumnack:2004,thesis:li:2004}.
Neutral current DIS events were selected by requiring that the scattered electron with energy more than 4 $\gev$ was measured in the 
CAL \cite{epj:c11:427}.

The offline kinematic variables $Q^2$ (four-momentum transfer squared), $x_{\rm Bj}$ (Bjorken scaling variable) and $y=Q^2/(sx_{\rm Bj})$ 
($s$ is the centre-of-mass energy squared)  were reconstructed by the electron ($e$), 
double angle (DA) \cite{proc:hera:1991:23}
and Jacquet-Blondel (JB) \cite{proc:epfacility:1979:391} methods. 
The angle of the hadronic system, $\gamma_{\rm had}$, corresponds, in the quark-parton model, to the direction of the scattered quark 
and was reconstructed from the CAL measurements of the hadronic final state.

If $\gamma_{\rm had}$ was less than 90$^\circ$ and the scattered-electron track could be 
well reconstructed by the CTD, the DA method was used; otherwise, the electron method was used.
The offline selection of DIS events was similar to that used in a previous ZEUS measurement \cite{epj:c23:13}
and was based on the following requirements:

\begin{itemize}
\item $E_e^\prime \gre 10\gev$, where $E_e^\prime$ is the scattered-electron energy after correction for energy loss in inactive material in front 
of the CAL, to achieve a high-purity sample of DIS events;

\item $y_e \les 0.6$, where $y_e$ is $y$ reconstructed by the electron method, to reduce the photoproduction background;

\item $y_{\rm JB} \gre 0.04$, where $y_{\rm JB}$ is $y$ reconstructed by the JB
method, to ensure sufficient accuracy for the DA reconstruction of $Q^2$;

\item $\cos\gamma_{\rm had} \les 0.7$, to ensure good reconstruction of jets in the Breit frame;

\item $40 \les \sum_i(E-P_Z)_i \les 60\gev$, where the sum runs over all CAL energy deposits. The lower cut removed background from
photoproduction and events with large initial-state QED radiation. The higher cut removed cosmic-ray background;

\item $|Z_{\rm vertex}| \les 50$~cm, where $Z_{\rm vertex}$ is the
reconstructed primary vertex $Z$-position, to select events consistent with $ep$
collisions; 

\item $|X| \gre 13$ or $|Y| \gre 7~{\rm cm}$, where $X$ and $Y$
are the impact positions of the scattered electron on the RCAL, to 
avoid the low-acceptance region adjacent to the rear beampipe.

\end{itemize}

The kinematic range of the analysis is defined as:

\begin{center}
$10<Q^2<5000\gev^2$ and $0.04<y<0.6$.
\end{center}
	
Jets were reconstructed using the $k_T$ cluster algorithm \cite{np:b406:187} in the 
longitudinally invariant inclusive mode \cite{pr:d48:3160}. 
The jet search was conducted in the Breit frame \cite{feynman:1972:photon,*zfp:c2:237}. 
For each event, the jet search was performed using a combination of track and CAL information,
excluding the cells and the track associated with the scattered electron. The selected tracks and CAL
clusters were treated as massless Energy Flow Objects (EFOs) \cite{desy-04-053}. The clustering
of objects was done according to the Snowmass convention\cite{proc:snowmass:1990:134}. 

The jet phase space is defined by selection cuts on the jet pseudorapidity
$\eta_{\rm LAB}^{\rm jet}$ in the laboratory frame and on the jet transverse energy
$E_{T,B}^{\rm jet}$ in the Breit frame:
\begin{center}
$-1 \les \eta_{\rm LAB}^{\rm jet} \les 2.5$ and $E_{T,B}^{\rm jet} \gre 5\gev$.
\end{center}
Events with two (three) or more jets were selected by requiring the invariant
mass of the two (three) highest $E_{T,B}^{\rm jet}$ jets to be:
\begin{center}
$M_{\rm 2jets (3jets)} \gre\ 25\gev$. 
\end{center}
These requirements were necessary to ensure a reliable prediction of the 
cross sections at NLO (see Section~\ref{nlo}). 

After all cuts, 37089 events with two or more jets (dijets) and
13665 events with three or more jets (trijets) remained.

\section{Monte Carlo simulation}
\label{mc}
Monte Carlo (MC) simulations were used to correct the data for detector 
effects, inefficiencies of the event selection and that of the jet reconstruction, as well as for QED effects. 
Neutral current DIS events were generated using the {\sc Ariadne}~4.08 program \cite{cpc:71:15} 
and the {\sc Lepto}~6.5 program \cite{cpc:101:108} interfaced to {\sc Heracles}~4.5.2 \cite{cpc:69:155} 
via {\sc Django}~6.2.4 \cite{cpc:81:381}. The {\sc Heracles} program includes
QED effects up to $\mathcal{O}(\alpha_{\rm EM}^{2})$.
In case of {\sc Ariadne}, the QCD cascade is 
simulated using the colour-dipole model \cite{np:b306:746}, whereas for {\sc Lepto}, the matrix 
elements plus parton shower model is used. Both models use the Lund string model \cite{prep:97:31}, 
as implemented in {\sc Jetset} 7.4 \cite{cpc:46:43,cpc:82:74}, for hadronisation. 

The ZEUS detector response was simulated with a program based on {\sc Geant} 3.13
\cite{tech:cern-dd-ee-84-1}. The generated events were passed through the 
detector simulation, subjected to the same trigger requirements as the data,
and processed by the same reconstruction and offline programs.

Measured distributions of kinematic variables are well described by both the {\sc Ariadne} 
and {\sc Lepto} MC models after reweighting in $Q^2$.
The {\sc Lepto} simulation gives a better overall description of the $E_{T,B}^{\rm jet}$
and invariant mass distributions. Therefore, the events generated with the {\sc Lepto}
program were used to determine the acceptance corrections. The events generated with
{\sc Ariadne} were used to estimate the systematic uncertainty associated with the treatment of
the parton shower.

\section{NLO QCD calculations}
\label{nlo}
The NLO calculations were carried out in the \MSbar scheme for
five massless quark flavors with the program {\sc Nlojet} \cite{prl:87:082001} using 
CTEQ6 \cite{hep-ph-0201195}, CTEQ4 \cite{pr:d55:1280}, MRST99 \cite{epj:c14:133} and \mbox{ZEUS-S \cite{pr:d67:012007}}
for the proton parton density functions (PDFs). {\sc Nlojet} allows a computation of
the trijet production cross sections to next-to-leading order, i.e. including all
terms up to $\mathcal{O}(\alpha_s^{3})$.
It was checked that the LO and NLO calculations from {\sc Nlojet} agree with those 
of {\sc Disent} \cite{np:b485:291} at the 1-2\% level for the dijet cross sections \cite{thesis:krumnack:2004,thesis:li:2004}.

For comparison with the data, the CTEQ6 
parameterisations of the proton PDFs were used
 and the renormalisation and factorisation scales 
were both chosen to be $(\bar{E}_T^2+Q^2)/4$, where 
for dijets (trijets) $\bar{E}_T$ is the average $E_T$ of the two (three) highest $E_T$ jets in a given event.
The strong coupling constant 
was set to the value used in the CTEQ6 analysis,
$\alpha_s(M_Z)=0.1179$, and evolved according to the two-loop
solution of the renormalisation group equation. 

The NLO QCD predictions were corrected for hadronisation effects using a
bin-by-bin procedure. Hadronisation correction factors
were defined for each bin as the ratio of the parton- to 
hadron-level cross sections and were calculated using the {\sc Lepto} MC program.
The correction factors $C_{\rm had}$ were typically in the range $1.15-1.35$ for most of the phase space.
 
\section{Corrections and systematic uncertainties}
\label{correction}

The jet transverse energy was corrected for energy losses in the 
inactive material in front of the CAL using the samples of MC simulated events \cite{np:b700:3}.
The cross sections for jets of hadrons in bins of $Q^2$, $E_{T,B}^{\rm jet}$ and $\eta_{\rm LAB}^{\rm jet}$ were 
obtained by applying a bin-by-bin correction to the measured jet distributions using 
the {\sc Lepto} program. The corrections take into account the efficiency of the trigger, the selection criteria and 
the purity and efficiency of the jet reconstruction. Additional 
corrections for QED effects, $C_{\rm QED}$, 
calculated using {\sc Heracles}, were applied 
to the measured cross sections, $\sigma_{\rm Born}=\sigma_{\rm meas.} \cdot
C_{\rm QED}$.  

A detailed study of the sources contributing to the systematic uncertainties of the measurements was 
performed \cite{thesis:krumnack:2004, thesis:li:2004}. The main sources contributing to
the systematic uncertainties are listed below (typical values of the systematic uncertainties in the 
dijet cross section and cross-section ratio $R_{3/2}$ are indicated in parentheses):

\begin{itemize}

\item {\it jet-pseudorapidity cut} - a change of $\pm$0.1 (corresponding to the resolution) 
      in the $\eta_{\rm LAB}^{\rm jet}$ cuts imposed on the jets in the laboratory frame for 
      both data and MC simulated events (1\%,1\%);

\item {\it jet transverse energy and invariant mass cuts} -  $E_{T,B}^{\rm jet}$
and $M_{\rm 2jets} (M_{\rm 3jets})$ 
      were simultaneously varied by the corresponding resolution near the cuts for both data and MC simulated events. 
      Along with the previous systematic check, this takes into account the 
      effect of the remaining differences 
      between the data and the MC simulation (3\%,3\%);
   
\item {\it use of different parton shower model} - using {\sc Ariadne} instead of {\sc Lepto} 
      to evaluate the acceptance corrections (2\%,4\%);
   
\item {\it the absolute energy scale of the CAL} - varying $E_{T,B}^{\rm jet}$ by its uncertainty of 
      $\pm1\% (\gre 10\gev)$ and $\pm3\% (\les 10\gev)$ for MC events \cite{pl:b547:164} (6\%,3.5\%).
\end{itemize}

The systematic uncertainties not associated with the absolute energy scale of the CAL were
added in quadrature to the statistical uncertainties and are shown on the figures as error bars.
The uncertainty due to the absolute energy scale of the CAL is highly correlated from bin-to-bin 
and is shown separately as a shaded band. The total systematic uncertainty and the uncertainty 
due to the absolute energy scale are also shown in Tables~\ref{et1}$-$\ref{rq2}.

The main contributions to the theoretical uncertainties of the NLO QCD predictions are:
\begin{itemize}

 \item {\it uncertainties in the proton PDFs}, which were estimated by repeating the calculations
 using 40 additional sets obtained under different theoretical assumptions as part of the CTEQ6 release (2.5\%,2\%);
 
 \item {\it uncertainties in the correction factors, $C_{had}$}, which were estimated by using the 
      {\sc Ariadne} program instead of {\sc Lepto} (6\%,4\%);
      
 \item {\it uncertainties due to terms beyond NLO}, which were estimated by varying
      both $\mu_R$ and $\mu_F$ between $(\bar{E}_T^2+Q^2)$ and $(\bar{E}_T^2+Q^2)/16$ (10\%,7\%).
 
\end{itemize}
The total theoretical uncertainty was obtained by adding in quadrature
the individual uncertainties listed above.

\section{Results}
\label{results}

\subsection{Differential cross sections}
\label{cross}

\begin{figure}[p]
\vfill
\begin{center}
\centerline{\epsfig{figure=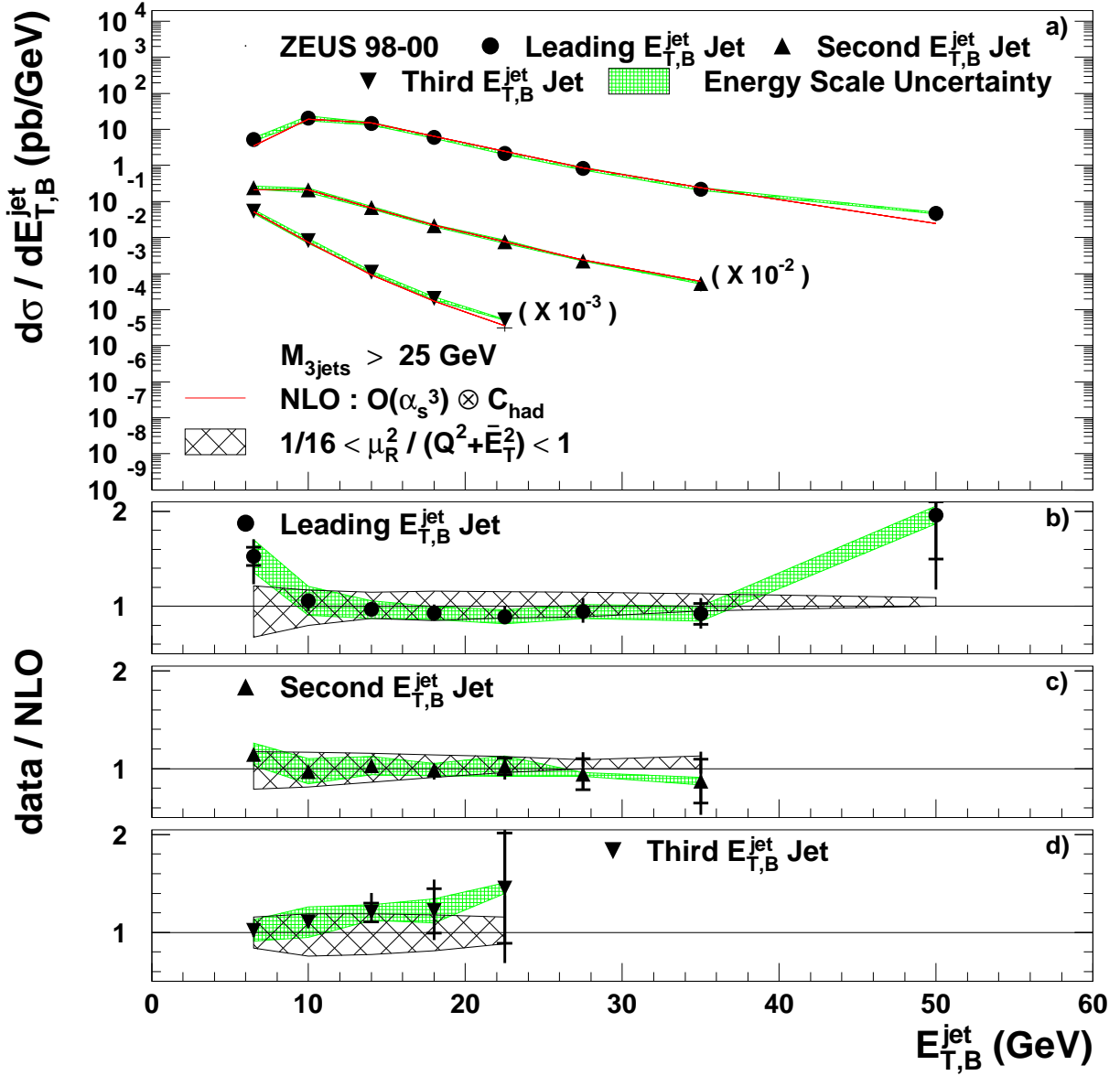,width=\linewidth}}
\end{center}
\caption{(a) The inclusive trijet cross sections as functions of
$E_{T,B}^{\rm jet}$ with the jets ordered in $E_{T,B}^{\rm jet}$. The cross sections of the
second and third jet were scaled by the factors shown for readability. 
The inner error bars
represent the statistical uncertainties. The outer error bars
represent the quadratic sum of statistical and systematic uncertainties
not associated with the calorimeter energy scale. The shaded band
indicates the calorimeter energy scale uncertainty. The predictions
of perturbative QCD in next-to-leading order, corrected for hadronisation
effects and using the CTEQ6 parameterisations of the proton PDFs, are compared to the data.
(b), (c) and (d) show the ratio of the data to the predictions. The hatched band
represents the renormalisation scale uncertainty of the QCD calculation.
}
\label{fig-e3}
\vfill
\end{figure}

The differential trijet cross sections as functions of $E_{T,B}^{\rm jet}$ are presented in Fig.~\ref{fig-e3} 
and in Tables~\ref{et1}$-$\ref{et3}.
The three highest $E_{T,B}^{\rm jet}$ jets were ordered in 
$E_{T,B}^{\rm jet}$ ($E_{T,B}^{\rm jet,1} \gre E_{T,B}^{\rm jet,2} \gre E_{T,B}^{\rm jet,3}$).
The observed decrease of the cross section for the first jet towards small values of $E_{T,B}^{\rm jet}$ 
is caused by the $E_{T,B}^{\rm jet}$ ordering combined with the requirement that the second and third jet have $E_{T,B}^{\rm jet} \gre 5\gev$.
For the second jet, a similar but less pronounced effect is observed. The NLO predictions using {\sc Nlojet}, 
corrected for hadronisation effects, are compared to the data in Fig.~~\ref{fig-e3}. The QCD predictions provide a good 
description of both the shape and magnitude of the measured cross sections, even at low $E_{T,B}^{\rm jet}$.

\begin{figure}[p]
\vfill
\begin{center}
\centerline{\epsfig{figure=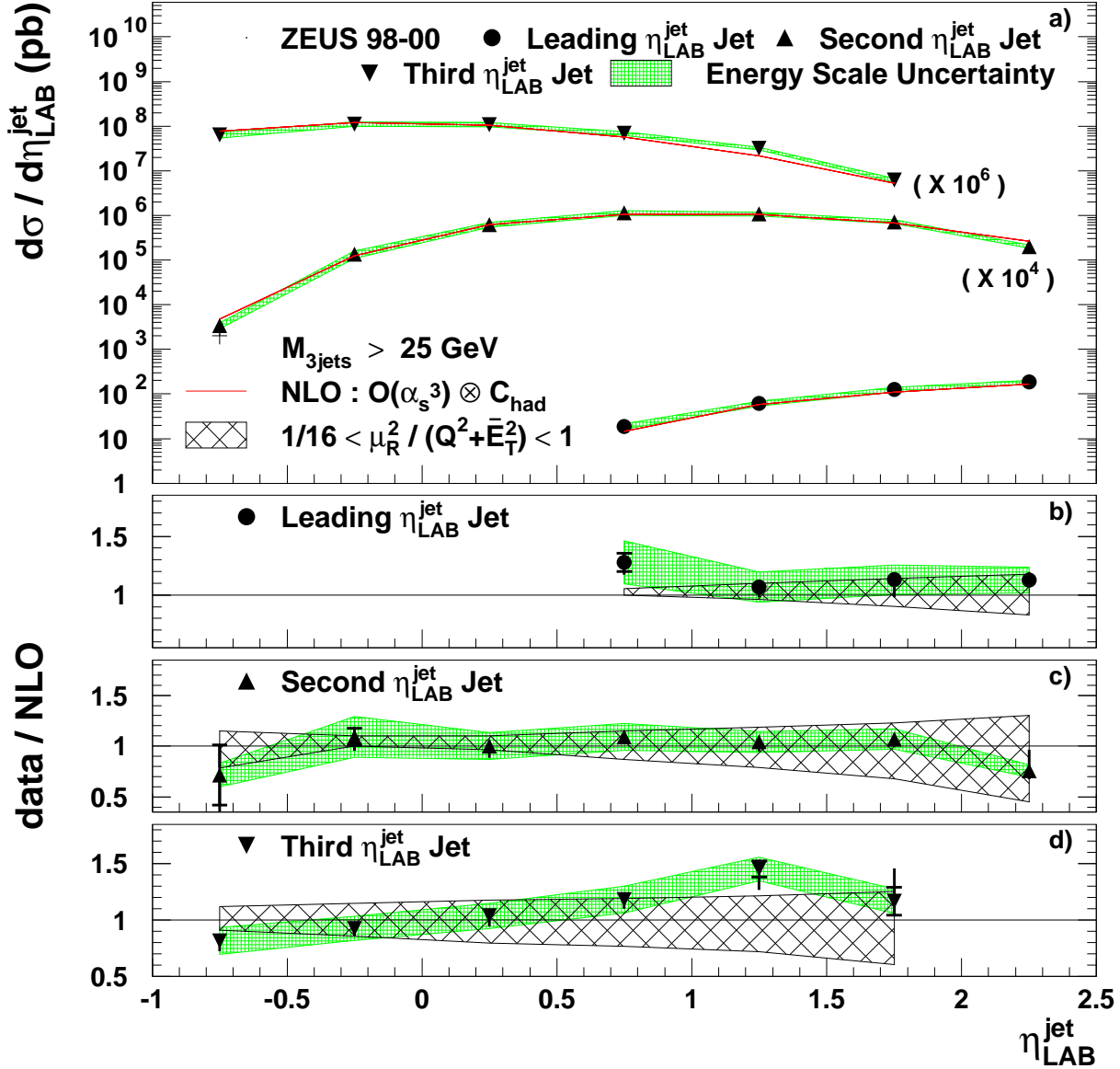,width=\linewidth}}
\end{center}
\caption{(a) The inclusive trijet cross sections as functions of
$\eta_{\rm LAB}^{\rm jet}$ with the jets ordered in $\eta_{\rm LAB}^{\rm jet}$. The cross sections of
the second and third jet were scaled up for readability only. 
The predictions of perturbative QCD in next-to-leading order are 
compared to the data. (b), (c) and (d) show the ratio of the data to the 
predictions. Other details are as in the caption to Fig.~\ref{fig-e3}.
}
\label{fig-h3}
\vfill
\end{figure}

\begin{figure}[p]
\vfill
\begin{center}
\centerline{\epsfig{figure=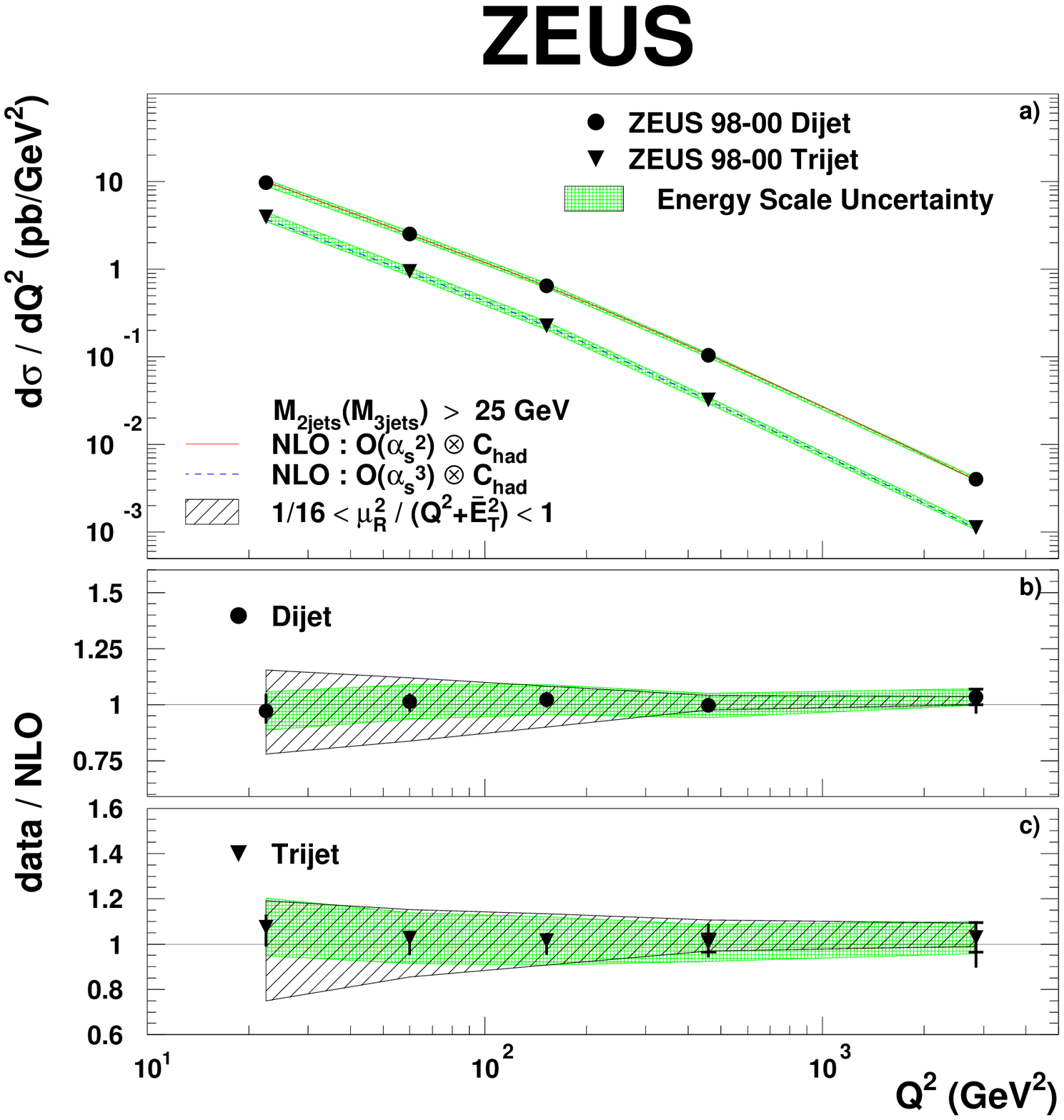,width=\linewidth}}
\end{center}
\caption{(a) The inclusive dijet and trijet cross sections as functions
of $Q^2$. The predictions of perturbative QCD in
next-to-leading order are compared to the data. (b) and (c) show the
ratio of the data to the predictions. Other details are as 
in the caption to Fig.~\ref{fig-e3}.
}
\label{fig-q1}
\vfill
\end{figure}

Figure~\ref{fig-h3} and Tables~\ref{eta1}$-$\ref{eta3} show the differential trijet cross sections 
as functions of $\eta_{\rm LAB}^{\rm jet}$. The three highest 
$E_{T,B}^{\rm jet}$ jets were ordered in $\eta_{\rm LAB}^{\rm jet}$ 
($\eta_{\rm LAB}^{\rm jet,1} \gre \eta_{\rm LAB}^{\rm jet,2} \gre \eta_{\rm LAB}^{\rm jet,3}$). 
Figure~\ref{fig-q1} and Tables~\ref{dq2} and ~\ref{tq2} show both the differential dijet and trijet cross section as functions of
$Q^2$. In Figs.~\ref{fig-h3} and \ref{fig-q1}, the data are generally well described by the NLO QCD predictions.
The largest difference is a slightly different slope of the $\eta_{\rm LAB}^{\rm jet}$ dependence of the third jet.

\subsection{Cross-section ratio and determination of $\alpha_s$}
\label{sec-alp}
\begin{figure}[p]
\vfill
\begin{center}
\centerline{\epsfig{figure=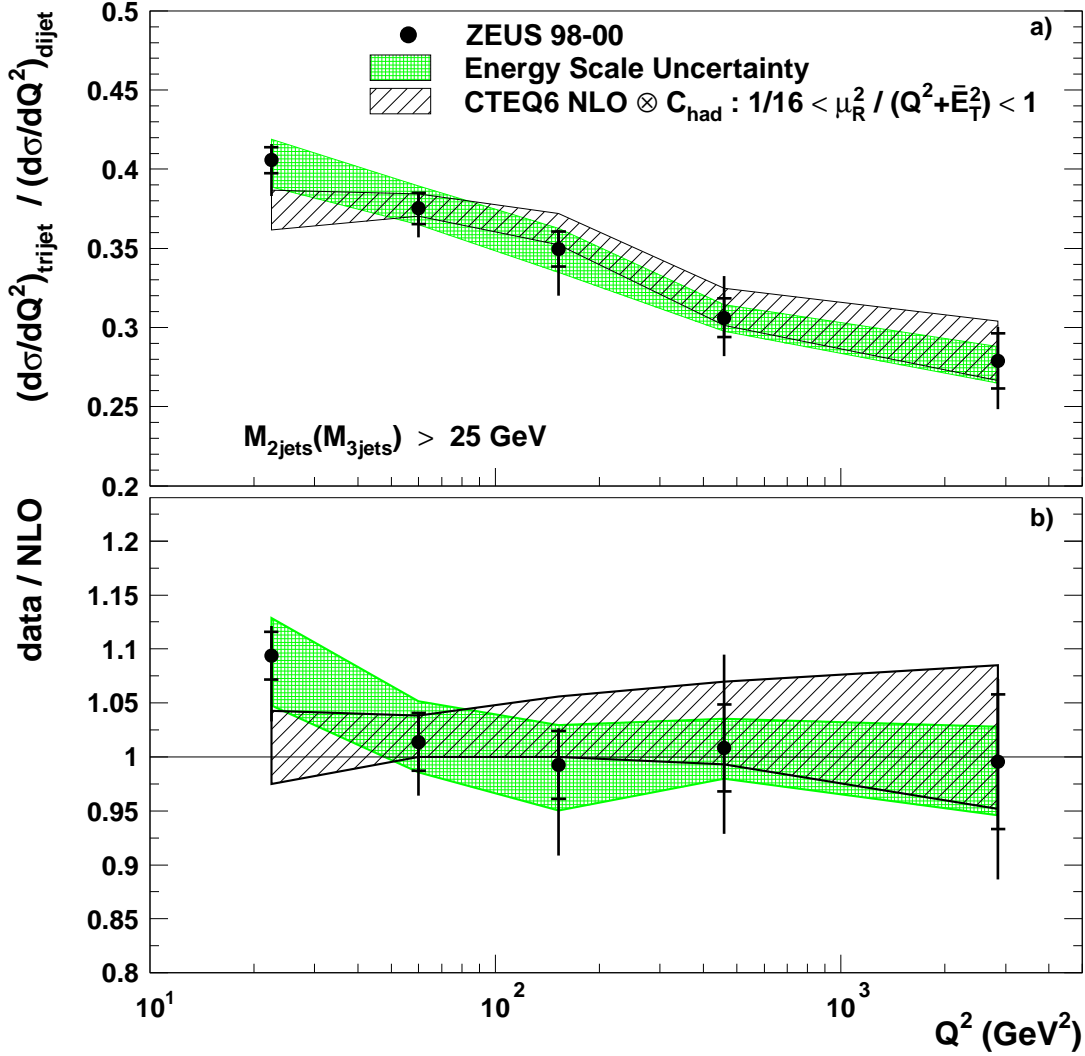,width=\linewidth}}
\end{center}
\caption{(a) The ratio of inclusive trijet to dijet cross sections as a
function of $Q^2$. The predictions of perturbative QCD in
next-to-leading order are compared to the data. (b) shows the ratio of the 
data to the predictions. 
Other details are as in the caption to Fig.~\ref{fig-e3}.
}
\label{fig-q2}
\vfill
\end{figure}

Figure~\ref{fig-q2} and Table~\ref{rq2} show the cross-section ratio $R_{3/2}$
of the trijet cross section to the dijet cross section,
as a function of $Q^2$. The correlated systematic and the renormalisation scale uncertainties largely cancel
in the ratio. The agreement between the data and NLO predictions is good. The
total experimental and theoretical
uncertainties are about $5\%$ and $7\%$, respectively. These uncertainties are substantially reduced with respect to those of
the di- and trijet cross sections. In particular, at low $Q^2$ ($Q^2 < 100\gev^2$), the theoretical uncertainties are reduced by as
much as a factor of four.
This reduction allows the determination of $\alpha_s(M_Z)$ at a much lower $Q^2$
than in
previous analyses ~\cite{np:b700:3,pl:b547:164}.

The measurement of $R_{3/2}$ as a function of $Q^2$ was used to determine $\alpha_s(M_Z)$ with
a method similar to that of a previous ZEUS publication \cite{pl:b507:70}:
\begin{itemize}

\item the NLO QCD calculation of $R_{3/2}$ was 
    performed using the five sets of proton PDFs of the CTEQ4
    A-series
    ~\cite{pr:d55:1280} {\footnote {The CTEQ4 PDF was chosen 
    because the 
    CTEQ6 does not provide PDF sets obtained with different 
    $\alpha_s(M_Z)$ values and therefore cannot be used for the 
    determination of $\alpha_s$.}}.
    The value of $\alpha_s(M_{Z})$ used in each partonic
    cross-section calculation was that associated with the
    corresponding set of PDFs: 0.110, 0.113, 0.116, 0.119, 0.122; 

\item for each bin, $i$, in $Q^2$, the NLO QCD calculations, corrected for 
hadronisation effects, were used to parameterise the $\alpha_s(M_Z)$ 
dependence of $R_{3/2}$ according to the functional form:

\begin{equation}
[R_{3/2}(\alpha_s(M_Z))]^i = C^i_1 \cdot \alpha_s(M_Z) + C^i_2 \cdot \alpha_s^2(M_Z),
\label{eqr}
\end{equation}

where $C^i_1$ and $C^i_2$ are fitting parameters. This simple parameterisation gives a good description of the $\alpha_s(M_Z)$ 
dependence of $R_{3/2}(Q^2)$ over the entire $\alpha_s$ range spanned by the PDF sets;

\item a value of $\alpha_s(M_Z)$ was then determined in each bin of $Q^2$, 
as well as in the entire $Q^2$ region,
by a $\chi^2$-fit of the measured $R_{3/2}(Q^2)$ values using the 
parameterisation in Eq.~(\ref{eqr}). 
\end{itemize}

This procedure correctly handles the complete $\alpha_s$-dependence of
the NLO differential cross sections (the explicit dependence coming from the
partonic cross sections and the implicit dependence coming from the PDFs) in the
fit, while preserving the correlation between $\alpha_s$ and the PDFs.
Taking into account only the statistical uncertainties on the measured cross-section ratio,
$\alpha_s(M_{Z})$ is determined to be $\alpha_s(M_Z)=0.1179\plm0.0013(\rm stat.)$.

\begin{figure}[p]
\begin{center}
\centerline{\epsfig{figure=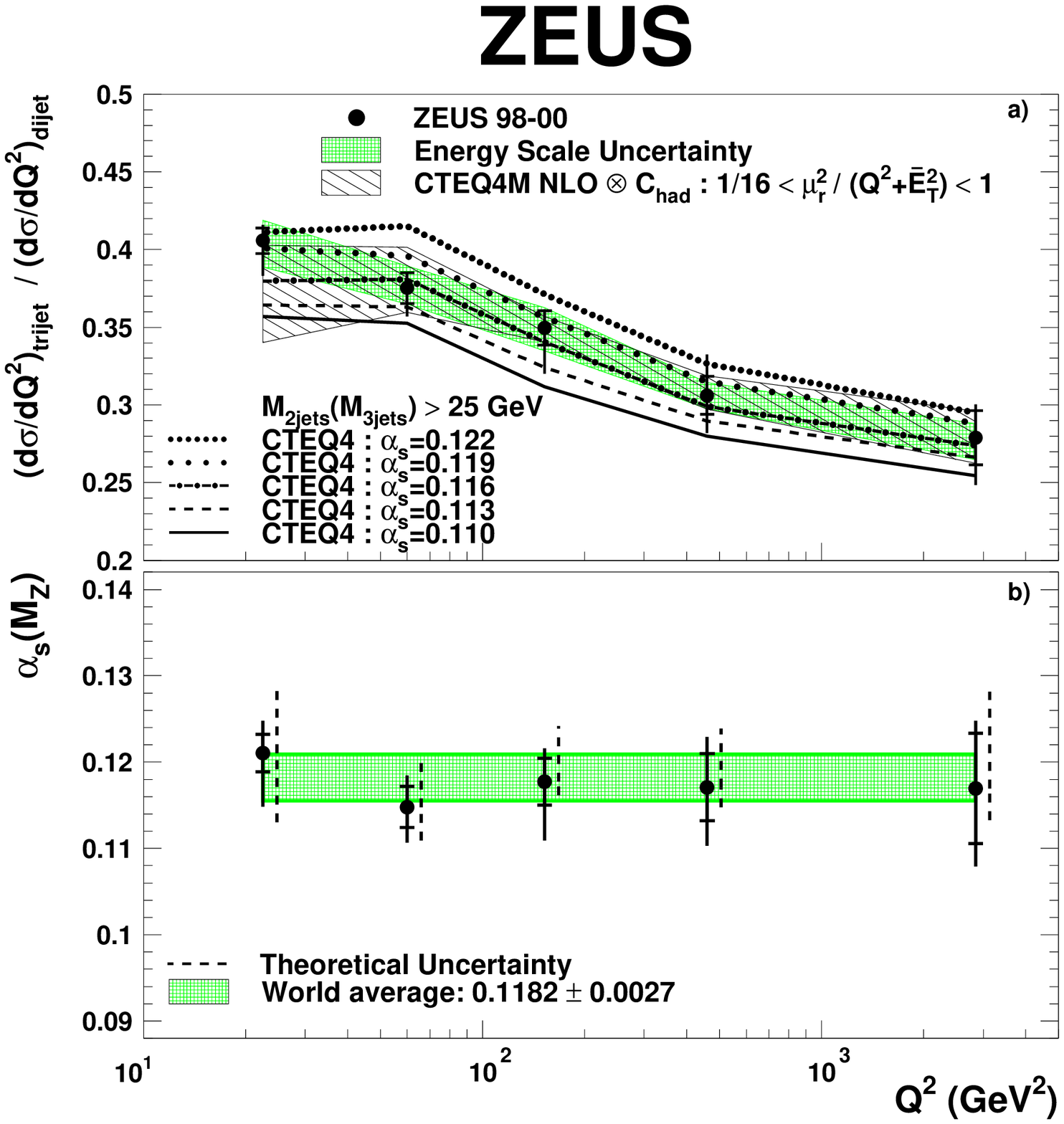,width=\linewidth}}
\end{center}
\caption{(a) The ratio of inclusive trijet to dijet cross sections as a
function of $Q^2$. The predictions of perturbative QCD in
next-to-leading order using five sets of CTEQ4 PDF are compared to the data.
(b) shows the $\alpha_s(M_Z)$ values determined from the 
ratio of inclusive trijet to dijet cross
sections in different regions of $Q^2$.
The shaded band indicates the current world average value of $\alpha_s(M_Z)$.
The inner error bars represent the statistical uncertainty of the data. 
The outer error bars show the statistical and systematic uncertainties added in quadrature.
The dashed error bars display the theoretical uncertainties. 
}
\label{fig-alphas}
\end{figure}

Figure~\ref{fig-alphas}a shows the sensitivity of the cross-section 
ratio $R_{3/2}$ to the value of $\alpha_s$. Figure~\ref{fig-alphas}b and 
Table~\ref{alphas} show $\alpha_s(M_Z)$
determined in the five bins of $Q^2$.

As a cross-check of the extracted value of $\alpha_s(M_Z)$, the fit procedure was
repeated by using the three sets of the MRST99 PDF corresponding to 
$\alpha_s (M_Z)$ equal to 0.1125, 0.1175 and 0.1225. 
The result is $\alpha_s(M_Z)=0.1178\plm0.0010(\rm stat.)$ 

In addition, the NLO QCD analysis used to obtain the ZEUS-S PDF ~\cite{pr:d67:012007}
was repeated to obtain a set
of five PDFs corresponding to the values of $\alpha_s(M_{Z})$: 0.115, 0.117,
0.119, 0.121, 0.123. 
These sets
were used in the current analysis yielding $\alpha_s(M_Z)=0.1191\plm0.0010(\rm stat.)$, in good agreement 
with the other determinations. 
 
The experimental and theoretical uncertainties of the extracted value of 
$\alpha_s(M_Z)$ were evaluated by repeating the analysis above for each systematic 
check, as described in Section~\ref{correction}. The main contributions to the
experimental systematic uncertainty were:
\begin{itemize}

\item {\it jet pseudorapidity cut} ($^{+1\%}_{-1.5\%}$);

\item {\it jet transverse energy and invariant mass cuts}  ($^{+0.5\%}_{-2\%}$);
   
\item {\it use of different parton shower model}  ($-2\%$) ;
   
\item {\it the absolute energy scale of the CAL}  ($^{+2\%}_{-2.5\%}$).
\end{itemize}

The main contributions to the theoretical uncertainty are: 
\begin{itemize}

 \item {\it uncertainties in the proton PDFs}  ($^{+1.5\%}_{-2\%}$);
 
 \item {\it uncertainties in the correction factor, $C_{\rm had}$}  ($+2\%$);
      
 \item {\it uncertainties due to terms beyond NLO} ($^{+5\%}_{-3.5\%}$).
 
\end{itemize}

The value of $\alpha_s(M_Z)$ as determined from the measurements of 
$R_{3/2}$ is therefore:

\begin{center}
$\alpha_s(M_Z) = 0.1179 \pm 0.0013~\rm(stat.)^{+0.0028}_{-0.0046}~\rm(exp.)^{+0.0064}_{-0.0046}~\rm(th.)$. 
\end{center}

The result is in good agreement with recent determinations at HERA 
\cite{pl:b363:201,pl:b507:70,pl:b547:164,pl:b558:41,pl:b560:7,np:b700:3,epj:c6:575,epj:c19:289,epj:c21:33} 
and the current world average of
$\alpha_s(M_Z)=0.1182\plm0.0027$~\cite{jp:g26:r27,*hep-ex-0407021}.

\section{Summary}
\label{sec-sum}
Differential dijet and trijet cross sections have been measured with high precision 
in neutral current deep inelastic scattering for $10<Q^2<5000\gev^2$ at HERA using the ZEUS detector. 
The inclusive trijet cross section has been measured as a
function of $E_{T,B}^{\rm jet}$, $\eta_{\rm LAB}^{\rm jet}$ and $Q^2$. The ratio $R_{3/2}$ of the
trijet and dijet cross sections has been measured
as a function of $Q^2$. The predictions of perturbative QCD calculations in
next-to-leading order give a good description of the dijet and trijet cross
sections and the cross-section ratio $R_{3/2}$ over the whole range of $Q^2$.
The cancellation of uncertainties in the ratio, in particular 
those from theory, allow the extraction of $\alpha_s$ with good
precision down to $Q^2$ of 10 GeV$^2$.
The value of the strong coupling constant $\alpha_s$ was measured to be 
$\alpha_s(M_Z) = 0.1179 \pm 0.0013~\rm(stat.)^{+0.0028}_{-0.0046}~\rm(exp.)^{+0.0064}_{-0.0046}~\rm(th.)$, 
in good agreement with the current world average value 
and previous determinations of $\alpha_s(M_Z)$ at HERA.

\section{Acknowlegements}
\label{sec-ack}
  
  We thank the DESY Directorate for their strong support and encouragement. The
  remarkable achievements of the HERA machine group were essential for the
  successful completion of this work and are greatly appreciated. 
  The design, construction and installation of the
  ZEUS detector has been made possible by the effort of many
  people who are not listed as authors.
  We would like
  to thank Z.~Nagy for useful discussions.

\vfill\eject

{
\def\bibname{\Large\bf References}
\def\refname{\Large\bf References}
\pagestyle{plain}
\ifzeusbst
  \bibliographystyle{./BiBTeX/bst/l4z_default}
\fi
\ifzdrftbst
  \bibliographystyle{./BiBTeX/bst/l4z_draft}
\fi
\ifzbstepj
  \bibliographystyle{./BiBTeX/bst/l4z_epj}
\fi
\ifzbstnp
  \bibliographystyle{./BiBTeX/bst/l4z_np}
\fi
\ifzbstpl
  \bibliographystyle{./BiBTeX/bst/l4z_pl}
\fi
{\raggedright
\bibliography{./BiBTeX/user/syn.bib,%
              ./BiBTeX/bib/l4z_articles.bib,%
              ./BiBTeX/bib/l4z_books.bib,%
              ./BiBTeX/bib/l4z_conferences.bib,%
              ./BiBTeX/bib/l4z_h1.bib,%
              ./BiBTeX/bib/l4z_misc.bib,%
              ./BiBTeX/bib/l4z_old.bib,%
              ./BiBTeX/bib/l4z_preprints.bib,%
              ./BiBTeX/bib/l4z_replaced.bib,%
              ./BiBTeX/bib/l4z_temporary.bib,%
              ./BiBTeX/bib/l4z_zeus.bib}}
}
\vfill\eject

\begin{table}[p]
\begin{center}
\begin{tabular}{||c|cccc||c||c||}
\hline
$E_{T,B}^{\rm jet,1}$ & $d\sigma/dE_{T,B}^{\rm jet,1}$ & 
$\delta_{\rm stat}$ & $\delta_{\rm syst}$ & $\delta_{\rm ES}$ & $C_{\rm QED}$ &
$C_{\rm had}$ \\
($\hbox{GeV}$) & ($\hbox{pb}/\hbox{GeV}$) & & & &  &  \\
\hline\hline
5 - 8 &  5.22 & $\pm$ 0.29 & $^{+0.75}_{-0.89}$ & $^{+0.63}_{-0.50}$ & 0.97 &
2.11 \\
8 - 12 & 20.2 & $\pm$ 0.5 & $^{+3.0}_{-3.5}$ & $^{+3.0}_{-2.9}$ & 0.96 & 1.33 \\ 
12 - 16 & 14.4 & $\pm $ 0.4 & $^{+1.6}_{-1.4}$ & $^{+1.6}_{-1.3}$ & 0.96 &
1.21 \\
16 - 20 & 6.05 & $\pm$ 0.22 & $^{+0.68}_{-0.54}$ & $^{+0.42}_{-0.44}$ & 0.96 &
1.19 \\
20 - 25 & 2.19 & $\pm$ 0.11 & $^{+0.32}_{-0.21}$ & $^{+0.20}_{-0.16}$ & 0.96 &
1.19 \\
25 - 30 & 0.828 & $\pm$ 0.068 & $^{+0.12}_{-0.10}$ & $^{+0.063}_{-0.065}$ & 0.97
& 1.22 \\
30 - 40 & 0.222 & $\pm$ 0.027 & $^{+0.036}_{-0.026}$ & $^{+0.021}_{-0.017}$ &
0.97 & 1.20 \\
40 - 60 & 0.047 & $\pm$ 0.012 & $^{+0.004}_{-0.012}$ & $^{+0.001}_{-0.004}$ &
1.07 & 1.31 \\
\hline
\end{tabular}
\caption{ The inclusive trijet cross section $d\sigma/dE_{T,B}^{\rm jet,1}$ for jets of
hadrons in the Breit frame, selected using the $k_T$ cluster algorithm in the 
longitudinally invariant inclusive mode. The statistical ($\delta_{\rm stat}$), 
systematic ($\delta_{\rm syst}$) and 
the absolute energy scale uncertainties ($\delta_{\rm ES}$) are shown separately. 
The multiplicative
correction factors for QED radiative effects ($C_{\rm QED}$), 
applied to the data, 
and for hadronisation
effects ($C_{\rm had}$), applied to the NLO predictions, are shown in the last two columns.}
\label{et1}
\end{center}
\end{table}

\begin{table}
\begin{center}
\begin{tabular}{||c|cccc||c||c||}
\hline
$E_{T,B}^{\rm jet,2}$ & $d\sigma/dE_{T,B}^{\rm jet,2}$ & 
$\delta_{\rm stat}$ & $\delta_{\rm syst}$ & $\delta_{\rm ES}$ & $C_{\rm QED}$ &
$C_{\rm had}$ \\
($\hbox{GeV}$) & ($\hbox{pb}/\hbox{GeV}$) & & & &  &  \\
\hline\hline
5 - 8 &  24.0 & $\pm$ 0.6 & $^{+2.6}_{-2.6}$ & $^{+2.5}_{-2.4}$ & 0.96 & 1.48 \\
8 - 12 & 20.7 & $\pm$ 0.4 & $^{+3.0}_{-2.8}$ & $^{+3.0}_{-2.6}$ & 0.97 & 1.22 \\ 
12 - 16 & 6.70 & $\pm $ 0.24 & $^{+0.65}_{-0.66}$ & $^{+0.65}_{-0.56}$ & 0.96 &
1.28 \\
16 - 20 & 2.16 & $\pm$ 0.13 & $^{+0.14}_{-0.22}$ & $^{+0.13}_{-0.14}$ & 0.95 &
1.32 \\
20 - 25 & 0.780 & $\pm$ 0.067 & $^{+0.087}_{-0.10}$ & $^{+0.086}_{-0.070}$ &
0.97 & 1.34 \\
25 - 30 & 0.225 & $\pm$ 0.038 & $^{+0.034}_{-0.005}$ & $^{+0.005}_{-0.005}$ &
0.97 & 1.40 \\
30 - 40 & 0.054 & $\pm$ 0.014 & $^{+0.013}_{-0.006}$ & $^{+0.004}_{-0.001}$ &
1.02 & 1.51 \\
\hline
\end{tabular}
\caption{ The inclusive trijet cross section $d\sigma/dE_{T,B}^{\rm jet,2}$ for jets of
hadrons in the Breit frame, selected using the $k_T$ cluster algorithm in the 
longitudinally invariant inclusive mode. Other details are as in the caption to 
Table~\ref{et1}.}
\label{et2}
\end{center}
\end{table}

\begin{table}
\begin{center}
\begin{tabular}{||c|cccc||c||c||}
\hline
$E_{T,B}^{\rm jet,3}$ & $d\sigma/dE_{T,B}^{\rm jet,3}$ & 
$\delta_{\rm stat}$ & $\delta_{\rm syst}$ & $\delta_{\rm ES}$ & $C_{\rm QED}$ &
$C_{\rm had}$ \\
($\hbox{GeV}$) & ($\hbox{pb}/\hbox{GeV}$) & & & &  &  \\
\hline\hline
5 - 8 &  52.8 & $\pm$ 0.8 & $^{+6.3}_{-6.2}$ & $^{+6.1}_{-5.3}$ & 0.96 & 1.28 \\
8 - 12 & 8.06 & $\pm$ 0.25 & $^{+1.1}_{-1.1}$ & $^{+1.1}_{-1.1}$ & 0.96 & 1.49 \\ 
12 - 16 & 1.11 & $\pm $ 0.09 & $^{+0.17}_{-0.08}$ & $^{+0.07}_{-0.08}$ & 0.97 &
1.66 \\
16 - 20 & 0.208 & $\pm$ 0.039 & $^{+0.025}_{-0.035}$ & $^{+0.025}_{-0.018}$ &
0.98 & 1.77 \\
20 - 25 & 0.052 & $\pm$ 0.020 & $^{+0.005}_{-0.019}$ & $^{+0.000}_{-0.003}$ &
0.94 & 1.77 \\
\hline
\end{tabular}
\caption{ The inclusive trijet cross section $d\sigma/dE_{T,B}^{\rm jet,3}$ for jets of
hadrons in the Breit frame, selected using the $k_T$ cluster algorithm in the 
longitudinally invariant inclusive mode. Other details are as in the caption to 
Table~\ref{et1}.}
\label{et3}
\end{center}
\end{table}

\begin{table}[p]
\begin{center}
\begin{tabular}{||c|cccc||c||c||}
\hline
$\eta_{\rm LAB}^{\rm jet,1}$ & $d\sigma/d\eta_{\rm LAB}^{\rm jet,1}$ & 
$\delta_{\rm stat}$ & $\delta_{\rm syst}$ & $\delta_{\rm ES}$ & $C_{\rm QED}$ &
$C_{\rm had}$ \\
 & ($\hbox{pb}$) & & & & &  \\
\hline\hline
0.5 - 1.0 & 19.1 & $\pm$ 1.2 & $^{+2.9}_{-2.7}$ & $^{+2.9}_{-2.6}$ & 0.95 &
2.18 \\
1.0 - 1.5 & 61.2 & $\pm$ 2.0 & $^{+7.3}_{-8.7}$ & $^{+7.3}_{-7.3}$ & 0.96 &
1.51 \\
1.5 - 2.0 & 125 & $\pm$ 3 & $^{+15}_{-20}$ & $^{+15}_{-13}$ & 0.96 & 1.31 \\
2.0 - 2.5 & 186 & $\pm$ 4 & $^{+23}_{-20}$ & $^{+20}_{-18}$ & 0.96 & 1.17 \\
\hline
\end{tabular}
\caption{ The inclusive trijet cross section $d\sigma/d\eta_{\rm LAB}^{\rm jet,1}$ for jets of
hadrons in the Breit frame, selected using the $k_T$ cluster algorithm in the 
longitudinally invariant inclusive mode. Other details are as in the caption to 
Table~\ref{et1}.}
\label{eta1}
\end{center}
\end{table}

\begin{table}[p]
\begin{center}
\begin{tabular}{||c|cccc||c||c||}
\hline
$\eta_{\rm LAB}^{\rm jet,2}$ & $d\sigma/d\eta_{\rm LAB}^{\rm jet,2}$ & 
$\delta_{\rm stat}$ & $\delta_{\rm syst}$ & $\delta_{\rm ES}$ & $C_{\rm QED}$ &
$C_{\rm had}$ \\
 & ($\hbox{pb}$) & & & & &  \\
\hline\hline
-1.0 - -0.5 & 0.34 & $\pm$ 0.14 & $^{+0.04}_{-0.18}$ & $^{+0.03}_{-0.09}$ &
0.91 & 8.72 \\
-0.5 - 0.0 & 13.6 & $\pm$ 1.0 & $^{+3.0}_{-2.4}$ & $^{+2.9}_{-2.1}$ & 0.98 &
2.30 \\
0.0 - 0.5 & 62.1 & $\pm$ 2.1 & $^{+9.3}_{-11.0}$ & $^{+9.1}_{-8.4}$ & 0.96 & 1.50 \\
0.5 - 1.0 & 115 & $\pm$ 3 & $^{+14}_{-14}$ & $^{+14}_{-13}$ & 0.97 & 1.29 \\
1.0 - 1.5 & 108 & $\pm$ 3 & $^{+12}_{-14}$ & $^{+11}_{-10}$ & 0.96 & 1.22 \\
1.5 - 2.0 & 73.3 & $\pm$ 2.2 & $^{+7.3}_{-7.8}$ & $^{+7.3}_{-6.3}$ & 0.96 &
1.19 \\
2.0 - 2.5 & 20.3 & $\pm$ 1.4 & $^{+5.6}_{-2.1}$ & $^{+1.7}_{-1.6}$ & 0.97 &
1.16 \\
\hline
\end{tabular}
\caption{ The inclusive trijet cross section $d\sigma/d\eta_{\rm LAB}^{\rm jet,2}$ for jets of
hadrons in the Breit frame, selected using the $k_T$ cluster algorithm in the 
longitudinally invariant inclusive mode. Other details are as in the caption to 
Table~\ref{et1}.}
\label{eta2}
\end{center}
\end{table}

\begin{table}[p]
\begin{center}
\begin{tabular}{||c|cccc||c||c||}
\hline
$\eta_{\rm LAB}^{\rm jet,3}$ & $d\sigma/d\eta_{\rm LAB}^{\rm jet,3}$ & 
$\delta_{\rm stat}$ & $\delta_{\rm syst}$ & $\delta_{\rm ES}$ & $C_{\rm QED}$ &
$C_{\rm had}$ \\
 & ($\hbox{pb}$) & & & & &  \\
\hline\hline
-1.0 - -0.5 & 63.7 & $\pm$ 2.3 & $^{+9.2}_{-11.0}$ & $^{+9.0}_{-9.3}$ & 0.96 &
1.32 \\
-0.5 - 0.0 & 112 & $\pm$ 3 & $^{+15}_{-14}$ & $^{+15}_{-12}$ & 0.96 & 1.24 \\
0.0 - 0.5 & 108 & $\pm$ 3 & $^{+12}_{-15}$ & $^{+12}_{-11}$ & 0.96 & 1.25 \\
0.5 - 1.0 & 68.6 & $\pm$ 2.1 & $^{+8.7}_{-7.6}$ & $^{+7.4}_{-6.5}$ & 0.97 & 1.32 \\
1.0 - 1.5 & 31.4 & $\pm$ 1.5 & $^{+2.3}_{-4.6}$ & $^{+2.1}_{-2.7}$ & 0.94 & 1.44 \\
1.5 - 2.0 & 6.21 & $\pm$ 0.67 & $^{+1.50}_{-0.67}$ & $^{+0.61}_{-0.61}$ & 0.93 &
1.50 \\
\hline
\end{tabular}
\caption{ The inclusive trijet cross section $d\sigma/d\eta_{\rm LAB}^{\rm jet,3}$ for jets of
hadrons in the Breit frame, selected using the $k_T$ cluster algorithm in the 
longitudinally invariant inclusive mode. Other details are as in the caption to 
Table~\ref{et1}.}
\label{eta3}
\end{center}
\end{table}

\begin{table}[p]
\begin{center}
\begin{tabular}{||c|cccc||c||c||}
\hline
$Q^2$  & $(d\sigma/dQ^2)_{\rm dijet}$ & 
$\delta_{\rm stat}$ & $\delta_{\rm syst}$ & $\delta_{\rm ES}$ & $C_{\rm QED}$ &
$C_{\rm had}$ \\
($\hbox{GeV}^2$) & ($\hbox{pb}/\hbox{GeV}^2$) & & & & &  \\
\hline\hline
10 - 35 &  9.70 & $\pm$ 0.10 & $^{+1.10}_{-0.94}$ & $^{+0.87}_{-0.76}$ & 0.97 &
1.18 \\
35 - 85 &  2.50 & $\pm$ 0.04 & $^{+0.21}_{-0.20}$ & $^{+0.20}_{-0.18}$ & 0.95
& 1.16 \\ 
85 - 220 & 0.649 & $\pm $ 0.011 & $^{+0.049}_{-0.044}$ & $^{+0.044}_{-0.041}$ &
0.96 & 1.15 \\
220 - 700 & 0.104 & $\pm$ 0.002 & $^{+0.006}_{-0.006}$ &
$^{+0.006}_{-0.006}$ & 0.94 & 1.12 \\
700 - 5000 & 0.00403 & $\pm$ 0.00014 & $^{+0.00015}_{-0.00024}$ &
$^{+0.00014}_{-0.00014}$ & 0.92 & 1.09 \\
\hline
\end{tabular}
\caption{ The inclusive dijet cross section $(d\sigma/dQ^2)_{\rm dijet}$ for jets of
hadrons in the Breit frame, selected using the $k_T$ cluster algorithm in the 
longitudinally invariant inclusive mode. Other details are as in the caption to 
Table~\ref{et1}.}
\label{dq2}
\end{center}
\end{table}

\begin{table}
\begin{center}
\begin{tabular}{||c|cccc||c||c||}
\hline
$Q^2$  & $(d\sigma/dQ^2)_{\rm trijet}$ & 
$\delta_{\rm stat}$ & $\delta_{\rm syst}$ & $\delta_{\rm ES}$ & $C_{\rm QED}$ &
$C_{\rm had}$ \\
($\hbox{GeV}^2$) & ($\hbox{pb}/\hbox{GeV}^2$) & & & & &  \\
\hline\hline
10 - 35 &  3.94 & $\pm$ 0.08 & $^{+0.54}_{-0.55}$ & $^{+0.49}_{-0.46}$ & 0.98 &
1.35 \\
35 - 85 &  0.94 & $\pm$ 0.02 & $^{+0.12}_{-0.11}$ & $^{+0.11}_{-0.09}$ & 0.95
& 1.31 \\ 
85 - 220 & 0.227 & $\pm $ 0.007 & $^{+0.024}_{-0.027}$ & $^{+0.024}_{-0.023}$
& 0.96 & 1.32 \\
220 - 700 & 0.0320 & $\pm$ 0.0013 & $^{+0.0036}_{-0.0031}$ &
$^{+0.0027}_{-0.0026}$ & 0.94 & 1.35 \\
700 - 5000 & 0.00112 & $\pm$ 0.00007 & $^{+0.00008}_{-0.00014}$ &
$^{+0.00008}_{-0.00009}$ & 0.92 & 1.33 \\
\hline
\end{tabular}
\caption{ The inclusive trijet cross section $(d\sigma/dQ^2)_{\rm trijet}$ for jets of
hadrons in the Breit frame, selected using the $k_T$ cluster algorithm in the 
longitudinally invariant inclusive mode. Other details are as in the caption to 
Table~\ref{et1}.}
\label{tq2}
\end{center}
\end{table}

\begin{table}
\begin{center}
\begin{tabular}{||c|cccc||c||c||}
\hline
$Q^2$  & $R_{3/2}$ & 
$\delta_{\rm stat}$ & $\delta_{\rm syst}$ & $\delta_{\rm ES}$ & $C_{\rm QED}$ &
$C_{\rm had}$ \\
($\hbox{GeV}^2$) & & & & & & \\
\hline\hline
10 - 35 &  0.406 & $\pm$ 0.008 & $^{+0.014}_{-0.027}$ & $^{+0.013}_{-0.017}$ &
1.01 & 1.14 \\
35 - 85 &  0.375 & $\pm$ 0.010 & $^{+0.014}_{-0.019}$ & $^{+0.014}_{-0.010}$ &
0.99 & 1.02 \\ 
85 - 220 & 0.350 & $\pm $ 0.011 & $^{+0.014}_{-0.031}$ & $^{+0.013}_{-0.015}$ &
1.00 & 1.15 \\
220 - 700 & 0.306 & $\pm$ 0.013 & $^{+0.024}_{-0.022}$ & $^{+0.008}_{-0.009}$
& 1.01 & 1.21 \\
700 - 5000 & 0.279 & $\pm$ 0.018 & $^{+0.016}_{-0.029}$ & $^{+0.009}_{-0.014}$ &
1.01 & 1.21 \\
\hline
\end{tabular}
\caption{ The ratio of inclusive trijet to dijet cross sections for jets of
hadrons in the Breit frame, selected using the $k_T$ cluster algorithm in the 
longitudinally invariant inclusive mode. Other details are as in the caption to 
Table~\ref{et1}.}
\label{rq2}
\end{center}
\end{table}

\begin{table}
\begin{center}
\begin{tabular}{||c|cccc||}
\hline
$Q^2$  & $\alpha_s(M_Z)$ & 
$\delta_{\rm stat}$ & $\delta_{\rm exp}$ & $\delta_{\rm theo}$ \\
($\hbox{GeV}^2$) & & & & \\
\hline\hline
10 - 35 &  0.1210 & $\pm$ 0.0022 & $^{+0.0031}_{-0.0058}$ & $^{+0.0074}_{-0.0080}$ \\
35 - 85 &  0.1148 & $\pm$ 0.0024 & $^{+0.0028}_{-0.0033}$ & $^{+0.0056}_{-0.0039}$ \\ 
85 - 220 & 0.1178 & $\pm$ 0.0027 & $^{+0.0027}_{-0.0063}$ & $^{+0.0064}_{-0.0016}$ \\
220 - 700 & 0.1171 & $\pm$ 0.0039 & $^{+0.0043}_{-0.0056}$ & $^{+0.0068}_{-0.0023}$ \\
700 - 5000 & 0.1170 & $\pm$ 0.0064 & $^{+0.0045}_{-0.0064}$ & $^{+0.011}_{-0.0037}$ \\
\hline
10 - 5000 & 0.1179 & $\pm$ 0.0013 & $^{+0.0028}_{-0.0046}$ & $^{+0.0064}_{-0.0046}$ \\
\hline
\end{tabular}
\caption{The $\alpha_s(M_Z)$ values as determined in this analysis.
The statistical ($\delta_{\rm stat}$), experimental ($\delta_{\rm exp}$) and 
theoretical ($\delta_{\rm theo}$) systematic uncertainties are shown separately.
}
\label{alphas}
\end{center}
\end{table}

%
%
\end{document}